\documentclass[twocolumn,showpacs,preprintnumbers,amsmath,amssymb]{revtex4}

\usepackage{graphicx}
\usepackage{dcolumn}
\usepackage{bm}

\begin{document}

\preprint{APS/123-QED}

\title{On accelerated Universe expansion}

\author{L. V. Verozub}

 \email{leonid.v.verozub@univer.kharkov.ua}
\affiliation{Kharkov National University\\
}

\date{\today}

\begin{abstract}
It is shown that   observed peculiarities of the Universe expansion are an inevitable consequence 
of the gravitational force properties following from   gauge-invariant gravitation  equations considered in
detail in  an author's paper in Annalen der Physik, v.17, 28 (2008).

\end{abstract}

\pacs{04.50+h; 98.80.-k}
\maketitle

Numerous data testifying that the most distant galaxies move away from us with
acceleration where  obtained  for the last 8 years \cite{riess}.  This fact poses serious problems  both for  fundamental
physics and astrophysics    \cite{weinberg}. In the present paper it is shown that the available  observational data are an inevitable consequence of properties of the gravitational force implying from   gauge-invariant  gravitation equations  \cite{Verozub08a}.
These equations  were tested successfully by binary pulsar $PSR~ 1913+16$    \cite{verkoch00}.

In Minkowski's space-time the radial component of the gravitational force of a point mass $M$  affecting  the
free- falling particle  of mass $m$ is $m \ddot{r}$ where the acceleration $\ddot{r}=d^{2}r/dt^{2}$ 
must be found from the gravitation equations in use.
According to \cite{Verozub08a} the force  is given by 
\begin{equation}
 F=-m\left[c^{2} C^{\prime}/2A +(A^{\prime}/2A  - 2 C^{\prime}/2C)  \dot{r}^{2}\right]  ,
\label{gravaccel1}%
\end{equation}
where
\begin{equation}
A=f^{\prime2}/C,\ C=1-r_\mathrm{g}/f,\ f=(r_\mathrm{g}^{3}+r^{3})^{1/3}, \, f^{\prime}=df/dt.
 \label{ABC}
\end{equation}
In this equation the dot denotes the derivative with respect to $t$, 
$r_\mathrm{g} =2GM/c^{2}$,  $G$ is the
gravitational constant, $c$ is the  speed of light at infinity, the prime denotes the
 derivative with respect to $r$. 

 For   particles at rest ($\dot{r} =0$ ) 
\begin{equation}
F=-\frac{GmM}{r^{2}} \left[
1-\frac{r_\mathrm{g}}{(r^{3}+r_\mathrm{g}^{3})^{1/3}} \right]
\label{ForceStat}
\end{equation}

\begin{figure}
\includegraphics[width=6cm]{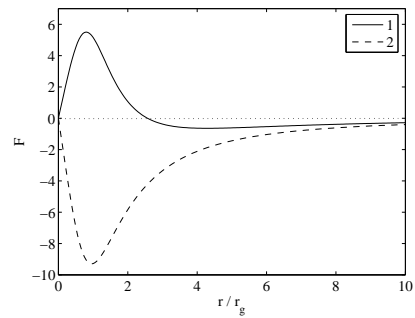}
 \caption{\label{fig: GRForce} The gravitational force (arbitrary units) affecting  a free-falling  particle  (the curve
1) and  a  particle  at rest (the curve 2) near an   attractive  point mass $M$.} 
\end{figure}

 Fig. \ref{fig: GRForce} shows the force $F$ 
affecting  a  particle at rest  and  a particle  free
falling from infinity    as the
function of the distance $\overline{r}=r/r_\mathrm{g}$ from the centre.

It follows from  Fig. \ref{fig: GRForce}  that the gravitational force affecting
free-falling particles  
 changes its sign at  $r \approx 2  r_\mathrm{g}$ . 
Although we have never yet observed particles motion at distances of the
order of $r_{g}$  we can verify this result for very distant objects in the
Universe, at large cosmological redshifts,  because it is well-known that the
radius of the observed region of the Universe is of the order of its
Schwarzschild radius.

A magnitude which is related with  observations in the expanding Universe
is the relative velocity of distant star objects with respect to an observer.
The radial velocity $v=\dot{R}=dR/dt$ of particles on the surface of a selfgravitating expanding 
homogeneous sphere of a radius $R$ 
 can be obtained from equations of the motion of a test particle 
 \cite{Verozub08a}:
\begin{equation}  \label{StarVelocity}
v=c\frac{C f^{2}}{R^{2}}\sqrt{1-\frac{C}{\overline{E}^{2}}},
\end{equation}
where $C$ are the functions of the distance $R$,  $r_{g}=(8/3)\pi c^{-2}G \rho R^{3}$ is 
Schwarzshild's radius   of  homogeneous matter inside of  the sphere 
 and    $\rho$ is the matter density. The parameter
  $\overline{E}$ is the constant total energy of a  particle divided by  $m c^{2}$.

\begin{figure}
\includegraphics[width=6cm]{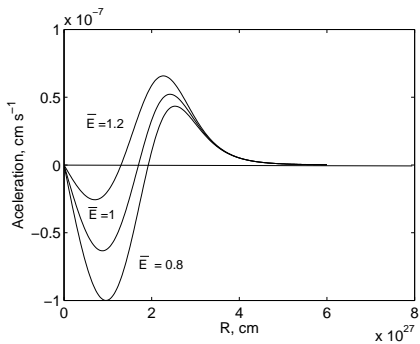}
\caption{\label{fig:aceleration} The acceleration of particles on the surface of an sphere of the
radius $R$ for three value of the parameter $\overline{E}$. The matter
density is equal to $10^{-29}{g}\,{cm}^{-3}$.}
\end{figure}

Fig. \ref{fig:aceleration} shows the radial acceleration $\ddot{R}=v'\dot{v}$ of 
a particle on the surface of the sphere of the radius $R$ in flat space-time 
Two conclusions can be made from this figure.\newline
\noindent 
1. At some distance from the observer the relative acceleration
changes its sign. If the $R < 2 \cdot 10^{27}{cm}$, the radial
acceleration of particles  is negative. If $R > 2 \cdot 10^{27}{cm}$,  the 
acceleration is positive. Hence, for
sufficiently large distances the  gravitational force affecting  particles 
is repulsive and gives rise to a relative radial acceleration of   particles with respect to any
observer. 

2. The gravitational force, affecting  the particles, tends to zero when $R$
tends to infinity.  (The same fact takes place as regards the force acting
on particles in the case of  static matter).   The reason of the fact is that
at a sufficiently large distance $R$  from the observer the Schwarzschild radius of
the matter inside of the sphere of the radius $R$ become of the same order as its radius.
Approximately at $R \sim 2\, r_{g}$ the gravitational force begin to decrease.
The ratio $R/r_{g}\rightarrow \infty$ tends to zero when $R$ tends to infinity, and under the
circumstances the gravitational force in the theory under consideration
tends to zero. 

 The above ball can be consedered as a part of the  flat  accelerating  Universe
because like the case of Einstein's equations a spherically-symmetric matter layer does not
create gravitational field inside itself  \cite{Verozub08a}.   Furthermore, to calculate the velocity of a particle at the distance $R$ from the observer there is no necessity to demand a global
spherical symmetry of matter outside of the sphere  due to  the above second properties of the gravitational force,  
because  the gravitational influence of very distant matter is neglected small. Therefore a relative 
velocity of   particles at the distance $R$ from the observer is determined by  gravitational field of the matter
inside of the sphere of the radius $R$. 

Proceeding from this equation we
 will find Hubble's diagram,  following mainly  the method being used in 
 \cite{zeldovich}.
 Let $\nu_{0}$.
be a  local frequency in the proper reference frame of a moving light source at the distance $R$ from an
observer, $\nu_{l}$ be this frequency in a local
inertial frame, and $\nu$ be the frequency as measured by the observer in the
sphere centre. The redshift $z=(\nu_{0}-\nu)/ \nu$ is caused by both 
Doppler effect and gravitational field. The Doppler effect is a consequence
of a difference between the local frequency of the source in inertial and
comoving reference frame, and it is given by  \cite{landau}

\begin{equation}
\label{DopplerRedshift}
\nu_{l}=\nu_{0}\, [1-\sqrt{(1-v/c)(1+v/c) } ] 
\end{equation}

The gravitational redshift is caused by the matter inside of the sphere of the
radius $R$.  
It is a consequence of the energy conservation for photon.  
According to equations of the motion of a test particle  \cite{Verozub08a}
the rest energy of a particle in gravitation field is given by  

\begin{equation}
E= mc^2 \sqrt{C}.
\end{equation} 
 Therefore, the difference in two local level  $E_{1}$ and $E_{2}$ of an atom energy in the field is $\Delta E=(E_{2}-E_{1}) \sqrt{C}$, so that the local frequency  $\nu_{0}$ at the distance $R$
from an observer are related with  the observed frequency  $\nu$ by equality
\begin{equation}
\label{redsiftGrav}
\nu=\nu_{l} \sqrt{C},
\end{equation}
where we take into account that for the observer location  $\sqrt{C}=1$.
 It follows from (\ref{redsiftGrav}) 
and (\ref{DopplerRedshift}) that the relationship between
frequency $\nu$ as  measured by the observer and the proper frequency $\nu_{0}$ 
of the moving source in the gravitational field takes the form 
\begin{equation}  \label{TotalRedshift}
\frac{\nu}{\nu_{0}} = \sqrt{C\, \frac{1-v/c}{1+v/c}}
\end{equation}

Equation (\ref{TotalRedshift}) 
yields the quantity $z$  as a function of $R$. 

By
solving this equation numerically  we obtain the
dependence $R=R(z)$   of the  measured distance  $R$   as a function of the redshift.
Therefore the distance modulus  \cite{weinberg2} to a star object is given by 
\begin{equation}  \label{muofz}
\mu=5\,  log_{10}[R(z)\, (z+1)] -5
\end{equation}
where $R(1+z)$ is a bolometric distance (in $pc$) to the object.

If  (\ref{StarVelocity}) is a correct equation for the radial relative velocity  of 
distant star objects  in the expansive Universe, it must  to  lead to the Hubble
law at small distances R. Under  this condition the Schwarzschild radius $%
r_{g}=(8/3) \pi G \rho R^{3}$ of the matter inside of  the sphere is very small
compared with $R$. For this reason $f\approx r$, and $C=1-r_{g}/r$. Therefore, at 
$\overline{E}=1$, we obtain from (\ref{StarVelocity}) that 
\begin{equation}
v=H R,  \label{Hubble}
\end{equation}
where 
\begin{equation}
\label{HubbleConstFormula}
H=\sqrt{(8/3) \pi G \rho}.
\end{equation}

If $\overline{E}\neq 1$ equation (\ref{StarVelocity}) does not lead to the
Hubble law,  since $v$ does not tend to zero when $R\rightarrow 0$. For this reason we set 
$\overline{E}= 1$ and look for the  value of the density at which a good
accordance with observation data can be obtained.

\begin{figure}[h ]
 \includegraphics[width=6cm]{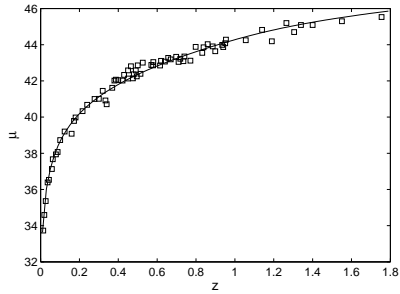}
\caption{\label{fig:HubbleFig}   The distance modulus $\protect\mu$ vs. the redshift $z$ for the density 
$\protect\rho=4.5 \cdot 10^{-30} g\, cm^{-3}$. Small squares denote the
observation data according to Riess et al. }
\end{figure}

The fig. (\ref{fig:HubbleFig}) show the Hubble diagram based on  eq. (\ref%
{muofz}) compared with observations data  \cite{riess}. 
It follows from this figure that the model under consideration are in a good accordance with  observation data. 

For  the value of the density $\rho=4.5 \cdot
10^{-30} g\, cm^{-3}$ we obtain from (\ref{HubbleConstFormula}) that 
\begin{equation}
H=1.59\cdot 10^{-18} c^{-1}=49\, km\, c^{-1}\, Mpc.
\end{equation}

\begin{figure}[h]
\includegraphics[width=6cm]{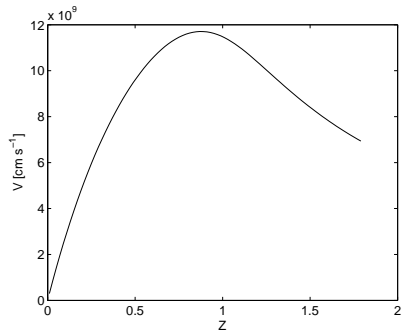}
\caption{\label{fig:VvsZ} The  radial velocity  vs. redshift $z$ for the density 
$\protect\rho=4.5 \cdot 10^{-30} g\, cm^{-3}$}
\end{figure}

Fig. \ref{fig:VvsZ} shows the dependence  of the radial velocity $v$ on the redshift. It
follows from this figure that  at $z> 1$  the Universe expands with an acceleration. 
 At $R\rightarrow \infty$ the velocity and acceleration tend to zero. 

\end{document}